\documentclass{aa}

\usepackage[varg]{txfonts}
\usepackage[english,english]{babel}
\usepackage[normalem]{ulem}
\usepackage[usenames, dvipsnames]{color}
\usepackage{graphicx}
\usepackage{natbib}
\usepackage{makecell}
\bibpunct{(}{)}{;}{a}{}{,}

\newcommand{\msun}{\mbox{$\rm M_\odot$}}

\newcommand{\br}{\mbox{$<b_r^2>$~}}
\newcommand{\fsz}{\mbox{$f_{\rm c}$~}}

\begin{document}

\title{AGN jets do not prevent the suppression of conduction by the heat buoyancy instability in simulated galaxy clusters}

\titlerunning{AGN jets and the HBI}
\authorrunning{Beckmann et al.}

\author{Ricarda S. Beckmann\inst{1,2}\thanks{ricarda.beckmann@iap.fr} \and Yohan Dubois\inst{1} \and Alisson Pellissier \inst{3} \and Fiorella L. Polles \inst{4} \and Valeria Olivares \inst{5}}
\institute{Institut d'Astrophysique de Paris, CNRS, Sorbonne Universit\'{e}, UMR7095, 98bis bd Arago, 75014 Paris, France
\and Institute of Astronomy and Kavli Institute for Cosmology, University of Cambridge, Madingley Road, Cambridge CB3 0HA, UK
\and AIM, CEA, CNRS, Université Paris-Saclay, Université Paris Diderot, Sorbonne Paris Cité, 91191 Gif-sur-Yvette, France
\and SOFIA Science Center, USRA, NASA Ames Research Center, M.S. N232-12 Moffett Field, CA 94035, USA
\and Department of Physics and Astronomy, University of Kentucky, 505 Rose Street, Lexington, KY 40506, USA
}

\date{Accepted . Received ; in original form }

\abstract{Centres of galaxy clusters must be efficiently reheated to avoid a cooling catastrophe. One potential reheating mechanism is anisotropic thermal conduction, which could transport thermal energy from intermediate radii to the cluster centre. However, if fields are not re-randomised, anisotropic thermal conduction drives the heat buoyancy instability (HBI) which reorients magnetic field lines and shuts off radial heat fluxes. We revisit the efficiency of thermal conduction under the influence of spin-driven AGN jets in idealised magneto-hydrodynamical simulations with anisotropic thermal conduction. Despite the black hole spin's ability to regularly re-orientate the jet so that the jet-induced turbulence is driven in a quasi-isotropic fashion, the HBI remains efficient outside the central 50 kpc of the cluster, where the reservoir of heat is the largest. As a result, conduction plays no significant role in regulating the cooling of the intra-cluster medium if central active galactic nuclei are the sole source of turbulence. Whistler-wave driven saturation of thermal conduction reduces the magnitude of the HBI but does not prevent it.}

\maketitle

%%%%%%%%%%%%%%%%%%%%%%%%%%%%%%%%%%%%%%%%%%%%%%%%%%

%%%%%%%%%%%%%%%%% BODY OF PAPER %%%%%%%%%%%%%%%%%%

\section{Introduction}

The hot intracluster medium (ICM) is a substantial thermal energy reservoir. If this energy can be efficiently transported to the cluster centre via thermal conduction, it could offset some of the centre's radiative cooling \citep{Fabian1994,Binney1995}, and thereby contribute to the long-term thermal stability of the cluster \citep{Tucker1983,DosSantos2001,Narayan01,Ruszkowski02}. 

The conductive heat flux $\vec{F}_{\rm cond}$  takes the form \citep{Spitzer1953}
\begin{equation}
    \vec{F}_{\rm cond}=-\kappa_{\rm e}\vec{\hat{b}}(\vec{\hat{b}} \cdot \nabla) T_{\rm e}
	\label{eq:Fcond}
\end{equation}
where $\kappa_{\rm e}$ is the Spitzer conductivity for electrons, $\vec{\hat{b}}$ is the unit vector along the magnetic field and $T_{\rm e}$ is the electron temperature. Assuming a predominantly radial temperature gradient, as seen in galaxy clusters, 
\begin{equation}
\vec{F}_{\rm cond,r} = - \kappa_{\rm e} \fsz  \frac{\partial T_{\rm e}(r)}{\partial r} 
\end{equation}
where \fsz parametrizes the effective strength of the conductive heating in comparison to the Spitzer value. As thermal conductivity is high along magnetic field lines but effectively zero across them \citep{Spitzer1953}, the effective efficiency of thermal conduction, \fsz, depends strongly on  magnetic field morphology. It is equal to $\fsz = b_r^2$, where $ b_r = \vec{\hat{b}} \cdot \mathbf{\hat{r}}$ is the magnetic field unit vector in the radial direction. A tangled magnetic field has $\fsz=1/3$, as the magnetic field is equally likely to be oriented in each of the three dimensions, whereas a radial field has $\fsz=1$. The required values of \fsz to offset radiative cooling depends on the cluster \citep{Jacob2017a}. 

In the presence of thermal conduction, the Heat-Buoyancy instability (HBI)~\citep{Quataert2008,Parrish2009} can reorient the cluster magnetic field. It acts when the temperature increases with height ($\mathbf{g} \cdot \nabla T_{\rm e} < 0$, where $\mathbf{g}$ is the gravitational acceleration). Left unchecked, the HBI will rearrange magnetic fields in galaxy cluster centres to a tangential configuration, suppressing conductive heat fluxes~\citep{Parrish2009,Bogdanovic2009}. Turbulence can counteract the HBI and re-randomise the magnetic field~\citep{Ruszkowski2010}. 

Thermal conduction has the potential to reduce cluster cooling flows \citep{Ruszkowski2011} and the total energy required for Active Galactic Nuclei (AGN) to regulate cluster cooling flows \citep{Kannan2017,Barnes2019}, but its efficiency depends on the efficiency of the HBI. This, in turn, depends on the relative magnitude of turbulent and buoyant timescales~\citep{McCourt2011}. Volume-filling turbulence of $50-100 \rm \ km s^{-1}$ can suppressed the HBI and allow for  $\fsz = 0.5$ \citep{Ruszkowski2010}. This is a level of turbulence that can be delivered by a simplified AGN-based turbulence model \citep{Parrish2012}, but more recent simulations employing fixed direction AGN jets found that the HBI remains active outside the jet cone \citep{Yang2016,Su2019}, which limits thermal conduction.

If AGN are only able to re-randomise magnetic fields around the jet cone, the jet direction is a key variable that determines the ability of the central AGN to prevent the HBI. In this paper, we revisit whether an AGN jet can offset the HBI and allow for efficient thermal conduction using a more self-consistent treatment of jet direction. The paper is structured as follows: the simulation setup is laid out in Section \ref{sec:setup}, insights on the evolution of the HBI are presented in Section \ref{sec:HBI}, the cooling flow is analysed in Section \ref{sec:cooling_flow} and conclusions are summarised in Section \ref{sec:conclusions}. 

\section{Simulations}
\label{sec:setup}

All simulations presented in this paper are part of the same suite as those presented in \citet{Beckmann2022}. We briefly summarise the setup here but refer the reader to \citet{Beckmann2022} for further details.

In this paper we present a set of magneto-hydrodynamical (MHD) simulations of isolated galaxy clusters with and without thermal conduction. The cluster simulations are run with the adaptive mesh refinement code {\sc ramses}~\citep{Teyssier2002} that solves for the MHD equations with separate ion-electron temperatures~\citep{Dubois2016}, including the anisotropic conductive heat flux  $\vec{F}_{\rm cond}=-\kappa_{\rm e}\vec{\hat{b}}(\vec{\hat{b}} \cdot \nabla) T_{\rm e}$  (see Eq. (\ref{eq:Fcond})). This includes the Spitzer conductivity,
\begin{equation}
\kappa_{\rm e}=f_{\rm sat}\kappa_{\rm sp}=f_{\rm sat} n_{\rm e}k_{\rm B} D_{\rm cond}
\end{equation}
  where $n_{\rm e}$ is the electron number density, $k_{\rm B}$ the Boltzmann constant and $D_{\rm cond}$ the thermal diffusivity. The conductive flux saturates once the characteristic scale length of the gradient of temperature $\ell_{T_{\rm e}}=T_{\rm e}/\nabla T_{\rm e}$ is comparable or shorter than the mean free path of electrons $\lambda_{\rm e}$~\citep{Cowie1977}. We follow~\citet{Sarazin1986} and introduce an effective conductivity that approximates the solution in the unsaturated and saturated regime by
\begin{equation}
\label{eq:sat}
    f_{\rm sat}=\frac{1}{1+4.2 \lambda_{\rm e}/\ell_{T_{\rm e}}}\, .
\end{equation}
When taken into account, the whistler instability \citep[e.g.][]{Roberg2016,Komarov18} can reduce the saturation coefficient to
\begin{equation}
\label{eq:sat_whist}
    f_{\rm sat,whist}=\frac{1}{1+(4.2+\beta/3) \lambda_{\rm e}/\ell_{T_{\rm e}}}\, 
\end{equation}
 for high plasma-$\beta$. Our simulations use Eq. (\ref{eq:sat}), except for AGN\_whistler, which uses Eq. (\ref{eq:sat_whist}) (See Table \ref{tab:simulations}). 

The induction equation is solved with constrained transport~\citep{Teyssier2006}, which guarantees $\nabla \cdot \vec{B}=0$ at machine precision. The MHD system equations are solved using the MUSCL-Hancock scheme~\citep{Fromang2006}, a minmod total variation diminishing scheme, and the HLLD Riemann solver~\citep{Miyoshi2005}. The flux for the anisotropic thermal conduction is solved with an implicit method~\citep{Dubois2016,Dashyan2020} using a minmod slope limiter on the transverse component of the face-oriented flux~\citep{Sharma2007}. 

\subsection{Initial conditions and Refinement}
Clusters are initialised with cored NFW profiles \citep{Navarro1997} (gas and dark matter) with a total mass of $8 \times 10^{14} \ \rm M_{\odot}$, a core radius of $13 \rm \ kpc$, a concentration parameter of $c_{\rm 200} = 4.41$ \citep{Maccio2007} and a gas fraction of $0.103$ \citep{Andreon2017}. The gas is initialised in hydrostatic equilibrium and dark matter is modelled as a fixed background potential. A black hole sink particle of mass $1.65 \times 10^{10} \rm \ M_\odot$ \citep{Phipps2019} is placed at the centre of the box. The magnetic field is initialised in a tangled configuration on a characteristic length scale of $10 \rm \ kpc$. It is scaled with the initial gas density profile $\rho(r)$ as $B(\rm r) = B_0(\rho(r)/\rho_0)^{2/3}$ where $\rm B_0 = 20 \rm \ \mu G$ and $\rho_0$ is the central density of the cluster.

Radiative cooling is calculated according to  \citet{Sutherland1993} for temperatures above $T>10^4 \,\rm K$, with values extended below $10^4$ K using \citet{Rosen1995}. Metallicity is treated as a passive scalar advected with the flow. It is initialised as $Z=\min\left(0.45,\max\left(0.22,0.15 \left(\frac{r}{r_{200}}\right)^{-0.28}\right)\right)$ using limits from \citet{Leccardi2008} and \citet{Urban2017}. Star formation proceeds in cells with hydrogen number density $n_{\rm H} > 0.1 \, \rm  H\, cm^{-3}$ and temperature $T < 10^4\,\rm K$, at a stellar mass resolution of $m_* = 3.89\times 10^5\, \msun$. Stellar feedback includes type II supernovae only, using the energy-momentum model of \citet{Kimm2015} using an efficiency of $\eta_{\rm SN} = 0.2$ and a metal yield of 0.1

The black hole accrets according to the Bondi-Hoyle-Lyttleton accretion rate, limited to a maximum Eddington fraction of 0.01 \citep{Dubois2012}. The black hole spin is initialised at zero (spin parameter $a=0$), and evolves throughout the simulation assuming a magnetically arrested disc \cite{McKinney2012}. A fraction $\epsilon_{\rm MAD}(a)$ of the accreted mass is returned as AGN bipolar kinetic outflows aligned with the black hole spin axis, using the implementation from  \citet{Dubois2014spin,Dubois2021}. As this spin vector naturally evolves under the influence of chaotic cold accretion on the black hole \citep{Gaspari2013}, there is no need to add further jet precession \citep{Beckmann2019b}. In AGN\_cr, a fraction $f_{\rm cr} = 0.1$ of the jet energy is injected into cosmic rays \citep[see][for details]{Beckmann2022}. During AGN feedback, a passive scalar is injected at the jet base which is used for refinement only. It decays with a decay time of 10 Myr.

The cluster profile is truncated at the virial radius ($r_{200} = 1.9 \rm \ Mpc$), and embedded in a box of size 8.7 Mpc. Simulations were performed on a root grid of $64^3$, adaptively refined to a maximum resolution of $\Delta x= 531 \rm \ pc $ if any of the following criteria are met: 1) The gas mass in a cell exceeds $[27089,8713,4098,1621,461,152,59,12,12] \times 1.47 \times 10^6 \rm \ M_\odot$ (levels 6 to 14). 2) The cell is located within $4 \Delta x$ of the black hole. 3) The cell density of the passive scalar injected by the jet exceeds $\rho_{\rm scalar} / \rho_{\rm gas}> 10^{-4}$ and its gradient exceeds 10\%.

Simulation parameters are summarised in Table \ref{tab:simulations}.

\begin{table}
    \centering
    \setlength\tabcolsep{3pt}
    \begin{tabular}{l c c c c}
    \bf Simulation & \bf AGN & \bf conduction & \bf \fsz & saturation \\ \hline
    noAGN\_nocond    & no & no & - & -\\
    noAGN\_cond & no & anisotropic & from $\vec{\hat{b}}$ & $f_{\rm sat}$\\
    AGN\_nocond & jet & no & - & - \\
    AGN\_cond & jet & anisotropic & from $\vec{\hat{b}}$ & $f_{\rm sat}$\\
    \hline
    AGN\_third & jet & anisotropic & 1/3 & $f_{\rm sat}$ \\
    AGN\_iso & jet & isotropic & 1 & $f_{\rm sat}$\\ \hline
    AGN\_whistler & jet & anisotropic & from $\vec{\hat{b}}$ & $f_{\rm sat,whist}$\\ 
    AGN\_cr* &   \makecell{jet,  \\ $f_{\rm cr} = 0.1$} & anisotropic & from $\vec{\hat{b}}$ & $f_{\rm whist}$\\ \hline
    \\
    \end{tabular}
    \caption{Simulation parameters. \fsz is the fraction of Spitzer conductivity used in the conductive heating rate of the simulation. *Simulation AGN\_cr has cosmic rays, and is identical to CRc\_dsh\_weak from \citet{Beckmann2022}.}
    \label{tab:simulations}
\end{table}

\section{Results}
\label{sec:results}

\subsection{AGN jets and the HBI}
\label{sec:HBI}

\begin{figure*}
    \centering
    \begin{tabular}{cc}
        \includegraphics[width=0.43\textwidth]{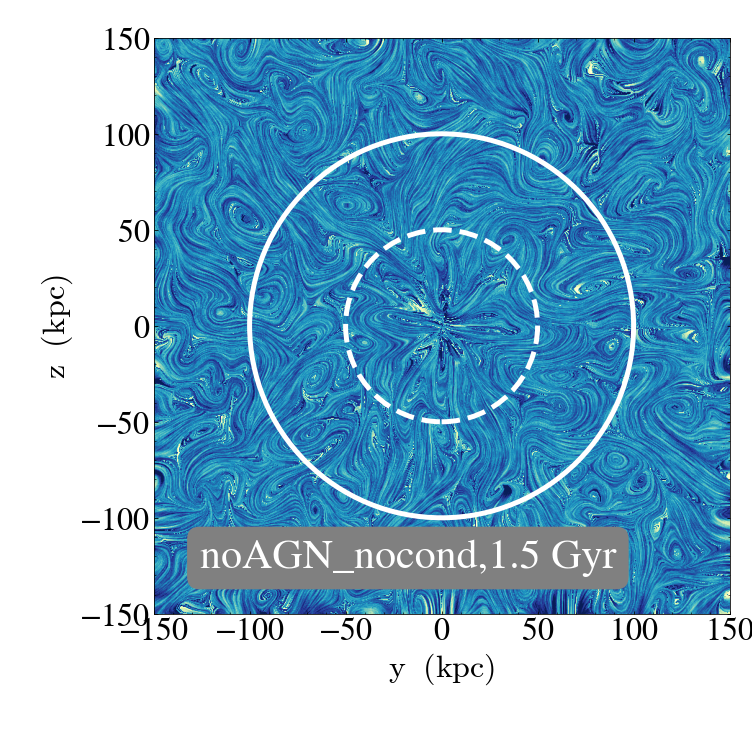} & 
        \includegraphics[width=0.43\textwidth]{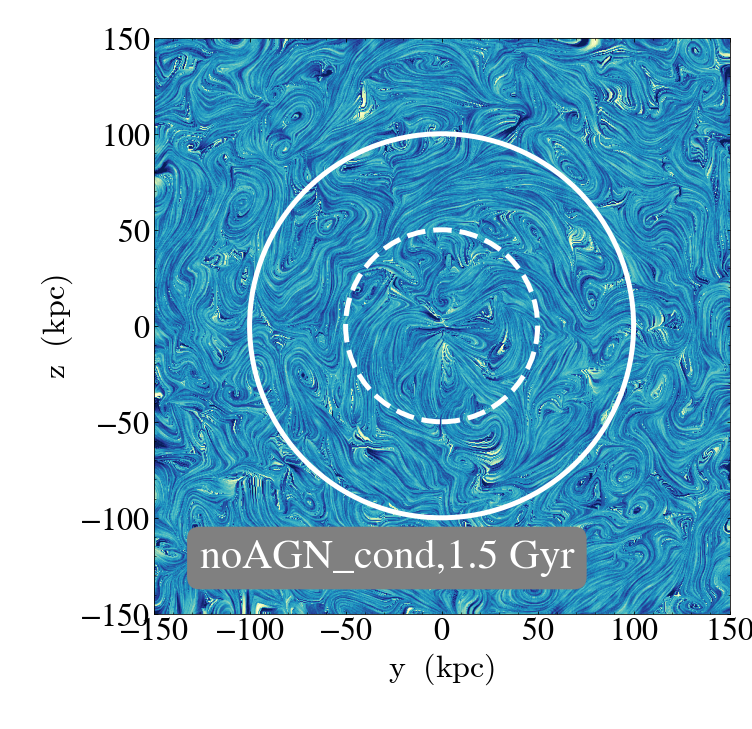} \\
        \includegraphics[width=0.43\textwidth]{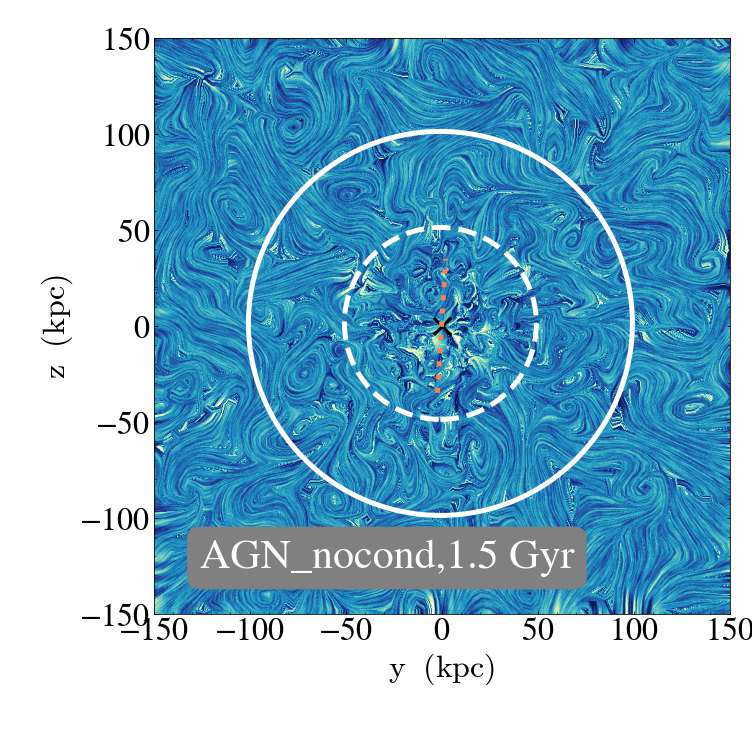} &
        \includegraphics[width=0.43\textwidth]{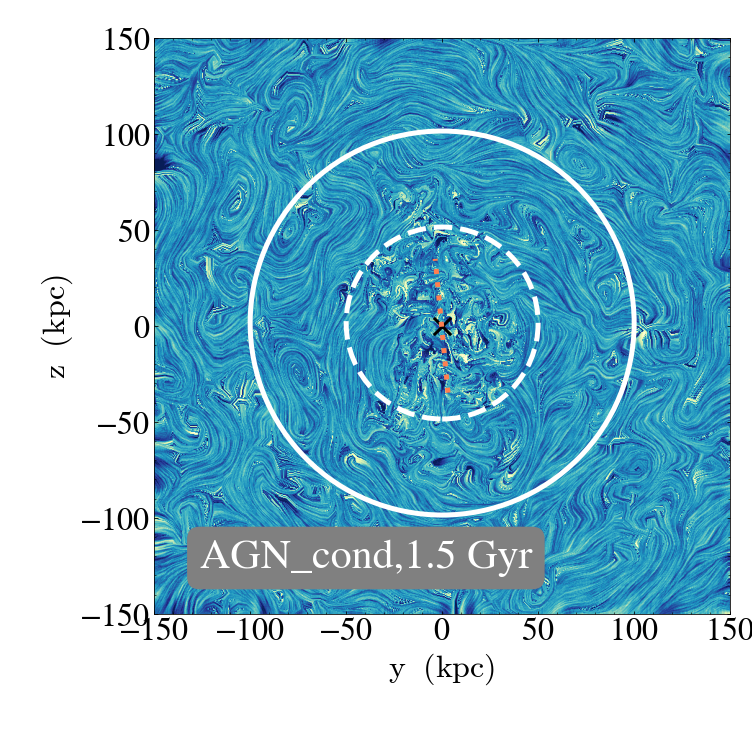} \\
        \includegraphics[width=0.43\textwidth]{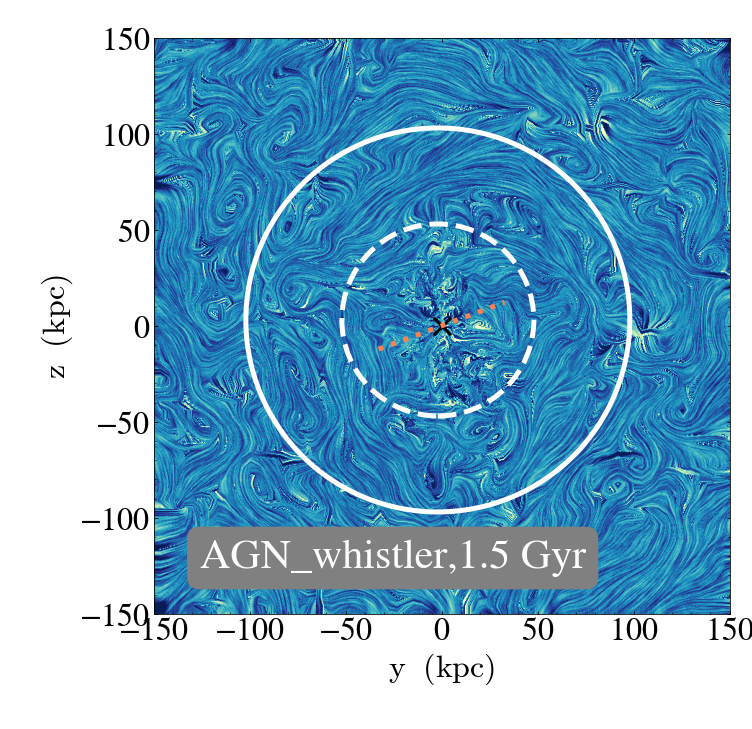} &
        \includegraphics[width=0.43\textwidth]{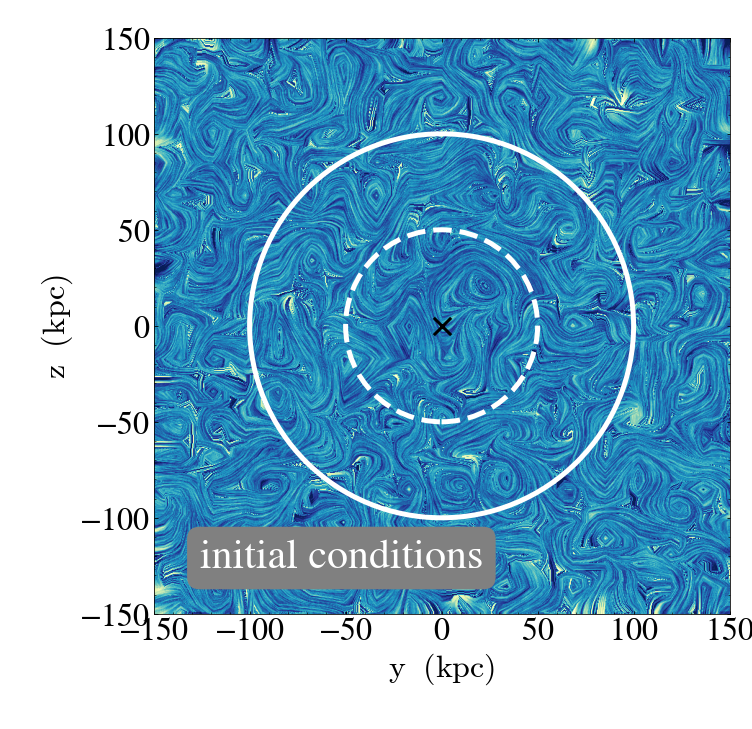} 
    \end{tabular}
    \caption{Line integral convolution of the magnetic field, in the plane of the image. Colourmaps are for visualization only. The black hole is marked as a black cross, and the instantaneous jet direction by an orange line. The white circle have radii of 50 (dashed) and 100 (solid) kpc respectively. In the presence of conduction, the HBI re-orients the magnetic field in a tangential configuration outside the central region.}
    \label{fig:images}
\end{figure*}

As can be seen in the magnetic field morphology shown in Fig. \ref{fig:images}, the presence of a self-regulating AGN only affects the magnetic field orientation in the central 50 kpc of the cluster. Without AGN, the central magnetic fields tend to be radial, while with AGN they are effective randomised by the jet and  resulting turbulence. At larger radii, conduction drives the HBI to tangentialize the magnetic field while in the absence of conduction, the magnetic field remains tangled (with an AGN) or develops radial features (without an AGN).

\begin{figure}
    \centering
    \includegraphics[width=\columnwidth]{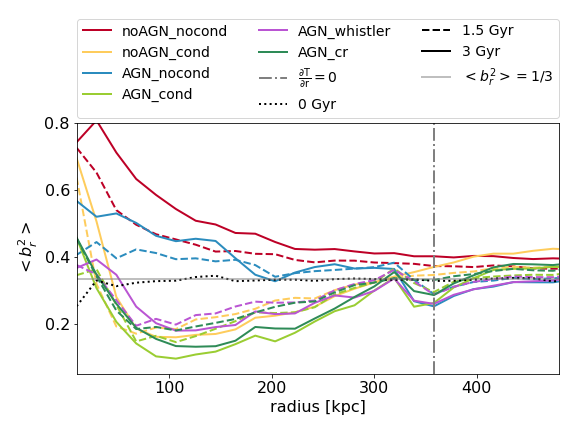}
    \caption{Radial profiles of \br for different simulations at two different points in time. The initial conditions for all simulations are shown as a black dotted line, and have on average \br$= 1/3$, as expected for a tangled magnetic field.}
    \label{fig:br_profiles}
\end{figure}

This evolution is shown quantitatively in Fig. \ref{fig:br_profiles} in the form of volume-weighted average radial profiles of the radial component of the magnetic field unit vector, \br, for concentric radial shells. For all simulations, the initial conditions (dotted black line) are consistent with a tangled magnetic field ($\br = 1/3$). From there, the evolution diverges: without conduction, \br increases (noAGN\_nocond and AGN\_nocond). With conduction (noAGN\_cond and AGN\_cond, AGN\_whistler and AGN\_cr), \br decreases out to the radius where the cluster temperature profile turns over (grey vertical line, $r= 358 \rm \ kpc$ at $t=0 \rm \ Gyr$), i.e. within the region where the HBI can act. 

\begin{figure}
    \centering
    \includegraphics[width=\columnwidth]{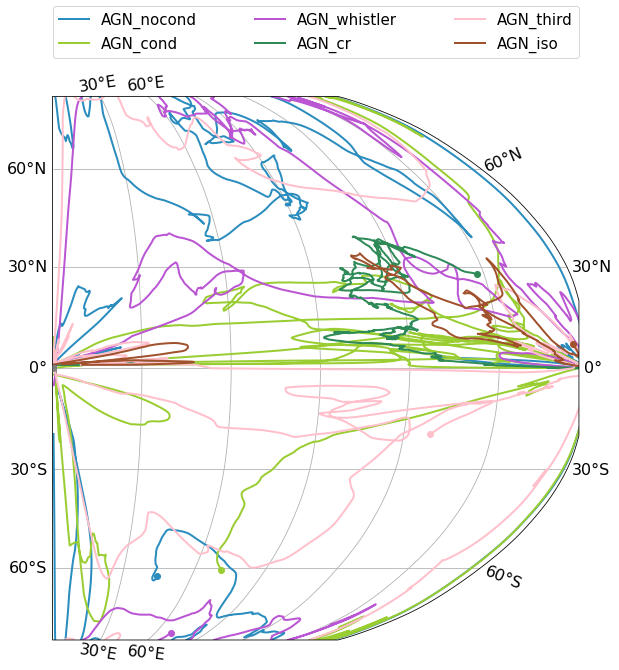}
    \caption{Evolution of the jet direction for all AGN simulations over 3 Gyr of evolution. Limits of the parameter space shown account for the symmetric nature of the jet. All jets show  significant reorientation over $3 \rm \ Gyr$ of evolution. A dot marks the jets final position at $3 \rm Gyr$.}
    \label{fig:spin}
\end{figure}

Low \br is a clear sign of the HBI in action, which in our simulation continues to decrease \br over the full $3 \rm \ Gyr$ of evolution. This is despite the fact that, unlike in \citet{Yang2016} and \citet{Su2019}, our jets sweep out a significant volume of the cluster centre over the $3 \rm \ Gyr$ evolution studied here. This can be seen in  Fig. \ref{fig:spin}, which shows the direction of jet axis  as a function of time. While this more isotropic injection of turbulence is effective at counteracting the HBI within the cluster centre, the limited extent of the jets meas values of \br still fall as low as 0.1 by $ r = 100 \rm \ kpc$ , which produces an effective barrier to heat fluxes to the cluster centre. In our model, the jet has a comparatively low average spin value of about $a=0.1$ \citep[see][for a discussion of the BH spin evolution]{Beckmann2019b}, which means the spin direction is more easily adjusted as for a faster spinning BH. It is possible that a jet with a more fixed direction would be able to inject turbulence to larger radii, but this would come at the cost of affecting a smaller fraction of the core volume. We will leave an investigation into the impact of the jet model on the evolution of the magnetic field morphology to future work.

Another possibility to stabilize the HBI would be via strong magnetic fields, as the HBI growth rate is damped on scales above $\rm H \gtrsim 20 \lambda_{\rm e} \beta $ \cite[][Eq. 33]{Quataert2008}, where $\beta$ is the plasma-$\beta$. In our clusters, the magnetic field strength after $3 \rm \, Gyr$ of evolution ranges from  $4-7.5 \, \mu \rm G$ in the centre to $0.5-1 \, \mu \rm G$ at $r=100 \rm \, kpc$. This means $H \geq \sim 640 \rm \, kpc$ in the centre and increases with radius, so the HBI is not suppressed by magnetic tension on scales relevant to the cluster cooling flow.

\begin{figure}
    \centering
    \includegraphics[width=\columnwidth]{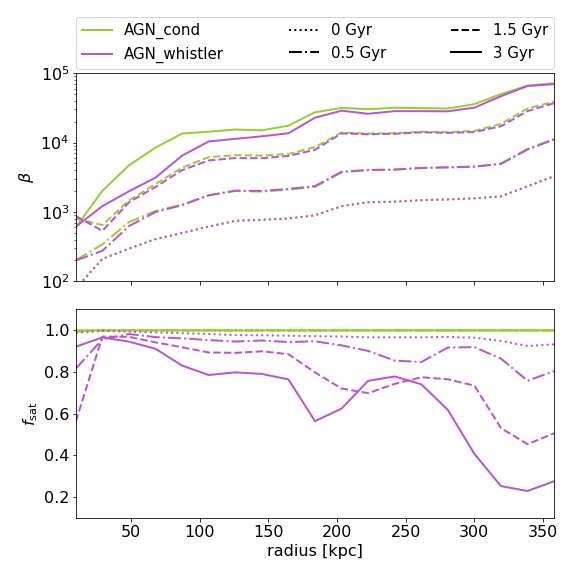}
    \caption{Volume-weighted radial profiles of the mean plasma beta (top) and saturation coefficient $f_{\rm sat}$ (bottom) for two the two choices of $f_{\rm sat}$ at three different points in time. At late times, conduction saturates at a lower value in the presence of whistler waves than in their absence. }
    \label{fig:whistler}
\end{figure}

Finally, whistler-wave modulated conduction could significantly delay the HBI. As can be seen in Fig. \ref{fig:br_profiles}, \br does indeed decrease more slowly for AGN\_whistler than for for AGN\_cond, but there is still a significant reduction in comparison to the random initial conditions. The presence of Whistler waves therefore slows down, but does not eliminate, the evolution of the HBI in galaxy clusters. This is due to the fact that for realistic magnetic field strengths of a few micro Gauss in the cluster centre, the plasma-$\beta$ in the cluster is sufficiently low such that $f_{\rm sat,whistler}\approx f_{\rm sat}$ early on, as can be seen in Fig. \ref{fig:whistler}. It is only as the magnetic field decays away slowly throughout the simulation due to the lack of volume-filling turbulence that plasma$-\beta$  increases and $f_{\rm sat, whistler} $ drops at late times.

The conclusion that Whistler-limited saturation makes only a small difference to the evolution of the HBI is in agreement with work by \citet{Berlok2021}, who studied the other conduction-driven instability, the magneto-thermal instability, which is active in cluster outskirts, and also concluded that whistler-based suppression of thermal conduction has only a small impact. We note that using a reduced saturation-coefficient is a simplified approach. A more complete treatment would be to self-consistently modelling the whistler-wave energy density, source and loss terms \citep{Drake2020}, and that doing so is likely to significantly change the conclusions on the impact of whistler waves on conduction and the HBI in galaxy clusters.

\br increases again at small radii for both noAGN\_cond and AGN\_cond but for different reasons: in AGN\_cond, the AGN randomises the magnetic field in the centre for an average $\br \sim 1/3$. In noAGN\_cond, a strong cooling flow develops, which causes radial inflows. Cooling flows are also responsible for the high values of noAGN\_nocond at all radii, and the late increase of \br for AGN\_nocond (See Section \ref{sec:cooling_flow}).

\subsection{Thermal conduction and the cooling flow}
\label{sec:cooling_flow}

\begin{figure}
    \centering
    \includegraphics[width=\columnwidth]{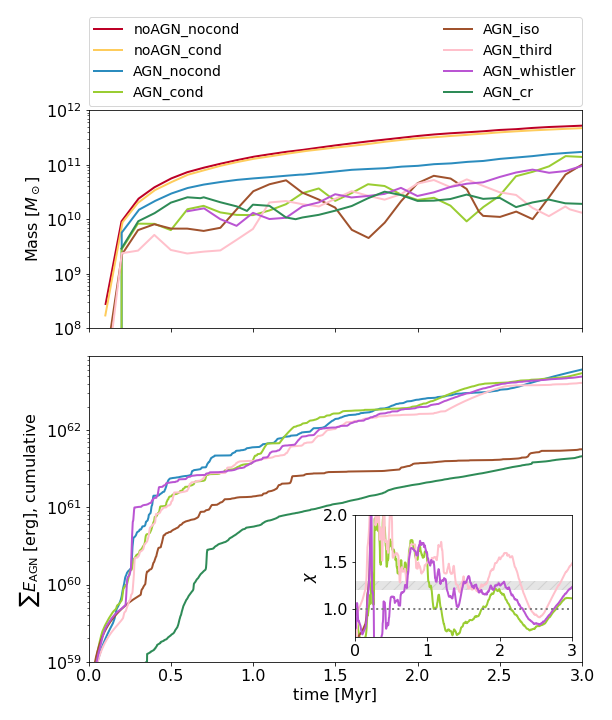}
    \caption{Total cold gas mass ($T<10^7 \rm \, K$, top) and cumulative AGN energy (bottom) as a function of time. The inset shows a timeseries of \mbox{$\chi = \sum \rm E_{\rm AGN, nocond} /  \sum \rm E_{\rm AGN, cond}$}, where $\sum \rm E_{\rm AGN,nocond}$ is the cumulative energy injected by the AGN up to time $t$ in the simulation without conduction, while $\sum \rm E_{\rm AGN,cond}$ is the same for simulations with different conduction prescriptions.    The grey shaded region reporting results from \citet{Kannan2017} and \citet{Barnes2019}. The presence of an AGN strongly reduces the cooling flow, but conduction only makes a significant difference when fully isotropic.}
    \label{fig:cooling_flow}
\end{figure}

The fact that both noAGN\_nocond and noAGN\_cond develop a strong cooling flow can be seen in Fig. \ref{fig:cooling_flow}. Both simulations without AGN build up unrealistically large quantities of cold gas in the cluster centre as the cooling of the hot ICM proceeds unimpeded. noAGN\_cond cools more slowly than noAGN\_nocond, but the reduction in total gas mass at 3 Gyr, $1.0 \times 10^{11} \rm \, M_\odot$ for noAGN\_cond versus $1.4 \times 10^{11} \rm \, M_\odot$ for noAGN\_nocond, is insignificant to the overall cooling flow. 

To further understand the potential impact of thermal conduction, we add three simulations with a fixed \fsz: noAGN\_iso and AGN\_iso  have $\fsz=1$, i.e. conduction is fully isotropic. AGN\_third also has isotropic conduction but uses $\fsz=1/3$, equivalent to a fully tangled magnetic field. When \fsz is constant, any reorientation of the magnetic field will have no impact on the heat flux. It is only when conduction is isotropic (noAGN\_iso), and therefore not influenced by the HBI, that conduction significantly reduces the cooling flow. This is in good agreement with work by \citet{Wagh2014}, who also showed that isotropic thermal conduction significantly reduces the formation of cold gas in galaxy clusters. However, even isotropic thermal conduction is less efficient at regulating the cooling flow than AGN feedback, as can be seen by comparing noAGN\_iso to any of the simulations with an AGN. 

In the presence of AGN jets, cooling flows are strongly reduced as cold gas building-up in the cluster activates the AGN which prevents further cooling \citep[see e.g.][for more details on AGN regulation of cluster cooling flows]{Li2015,Prasad2015a,Li2017,Yang2016,Beckmann2019b}. Comparing AGN\_nocond and AGN\_cond in Fig. \ref{fig:cooling_flow} shows that while the AGN effectively regulates cluster cooling and prevents a run-away cooling flow (top panel), anisotropic conduction makes little difference to the long-term evolution of the cluster. 

To compare the impact of thermal conduction on the ability of the AGN to self-regulate the cluster, we define \mbox{$\chi = \sum \rm E_{\rm AGN, nocond} /  \sum \rm E_{\rm AGN, cond}$} (inset, Fig. \ref{fig:cooling_flow}), where $ \sum \rm E_{\rm AGN}(t)$ is the cumulative energy injected by the AGN up to time $t$. In our simulations, $\chi \geq 1.2$, i.e. the simulation without conduction requires at least 20\% more cumulative AGN energy than the case with conduction, for all simulations only while $t< \rm \  1 \rm Gyr$, during which the cluster is still evolving away from the initial conditions. At late times, at an average value of $\chi = 1.24$ at $t>2 \rm \ Gyr$, $\chi$ remains elevated for AGN\_third,  but drops to $\sim 1$ for AGN\_cond and AGN\_whistler. This is in contrast to \citet{Kannan2017} and \citet{Barnes2019}, who report $\chi = 1.2 - 1.3$ at all times. One possibility is that the HBI in \citet{Kannan2017} and \citet{Barnes2019} is being offset by turbulence injected by the large-scale cosmological environment. Support for this theory comes from the fact that fully tangled fields (AGN\_third) in our simulations also show $\chi = 1.2 - 1.3$. However, \citet{Yang2016} used isolated clusters and also report $\chi \approx 1.5$. Another possibility is that as $\chi$ is very sensitive to the cluster mass \citep{Yang2016} so the different $\chi$ could be due to different cluster masses or cluster profiles. A final possibility is that the higher resolution in our simulation ($\Delta x = 531 \rm \ pc$ compared to $1.5 - 2 \rm \ kpc$ for other studies) is responsible for the higher $\fsz$  ( $\fsz \sim  0.2 $ in \citet{Yang2016} versus $\fsz \sim 0.1$ here) and the resulting difference in $\chi$.

Injecting a small fraction of AGN energy into cosmic rays (AGN\_cr) increases the efficiency of AGN jets in self-regulating clusters ($\rm \sum E_{\rm AGN}$ is reduced) but the cooling flow is only mildly affected. Further details on the impact of cosmic rays on galaxy cluster cooling flows can be found in \citet{Beckmann2022}. It is only when the magnetic field becomes preferentially radial ($\fsz \rightarrow 1$) that thermal conduction significantly reduces cold gas mass and cumulative AGN energy. This is in agreement with \citet{Jacob2017a} who also reported values of $\fsz > 1/3 $ for cosmic ray and thermal conduction regulated steady-state solutions. 

\begin{figure}
    \centering
    \includegraphics[width=\columnwidth]{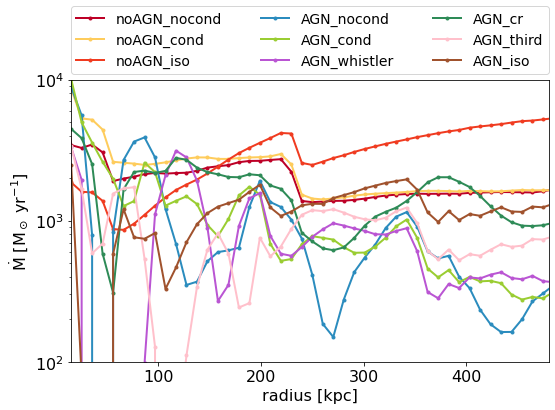}
    \caption{Net mass accretion rate through concentric radial shells averaged over $3 \rm \ Gyr$ of evolution in $500 \rm \ Myr$ bins. The simulations without AGN show steady, smooth inflows at all radii. Simulations with AGN show on average lower accretion rates and more disturbed accretion patterns due to the large-scale shocks propagating outwards from the central AGN.}
    \label{fig:mflux_inflow}
\end{figure}

Such cooling flows result in significant radial mass fluxes, as can be seen in Fig. \ref{fig:mflux_inflow}. Simulations without AGN show steady inflows at all radii, while those with AGN show lower net accretion rates with more disturbed patterns.The flow in simulations with AGN are less smooth as large-scale shocks propagating outwards from the central AGN disturb the contraction of the ICM as it looses thermal pressure support. In the very centre ($r< \rm 50 \ kpc$), inflows in AGN simulations are highly variable due to the turbulence and the multiphase nature of the gas~\citep{Beckmann2019b}.

It is radial mass flows that create the high values of \br for noAGN\_nocond in Fig. \ref{fig:br_profiles}. For ideal MHD, field-lines are "frozen in", i.e. advected with the gas flow. As material falls onto the cluster centre, these mass fluxes will drag the magnetic field in a radial configuration, which increases \br. This radialisation of magnetic field lines proceeds unimpeded without conduction (noAGN\_nocond) but competes with the tangentialisation due to the HBI in the presence of conduction. Comparing Fig. \ref{fig:mflux_inflow} with Fig. \ref{fig:br_profiles} shows that simulations with strong mass fluxes have on average higher \br than equivalent simulations without strong mass fluxes. For example, due to the absence of a strong cooling flow, \br is lower at 100 kpc in AGN\_cond than without AGN feedback (noAGN\_cond) despite the HBI being very active in both. Cooling flows also explain the late rise in \br for AGN\_nocond due to an excess of cold (Fig. \ref{fig:cooling_flow}), inflowing (Fig. \ref{fig:mflux_inflow}) gas at $t> 2.5 \rm \ Gyr$. Mass fluxes are high for noAGN\_iso because thermal conduction efficiently extracts thermal energy energy from intermediate radii which causes the gas to loses pressure support and contract.

\section{Conclusions}
\label{sec:conclusions}

In this paper, we investigated whether the presence of a black hole spin-driven AGN jet can counteract the HBI in the centre of massive galaxy clusters and allow for efficient thermal conduction to aid in the long-term self-regulation of cluster cooling flows. We conclude that
\begin{itemize}
    \item Spin-driven AGN feedback is able to randomize the magnetic field in the central $50 \rm \ kpc$ of the cluster but not outside this region.
    \item Whether an AGN is included or not, the HBI remains very active in the region 50 - 300 kpc from the cluster centre, which reduces the effective conductivity to values as low as $\fsz = 0.1$ of Spitzer conductivity.
    \item Such low levels of thermal conduction have no significant influence on the cluster cooling flow, or the AGN self-regulation thereof. 
    \item Whistler-wave driven saturation of thermal conduction reduces the magnitude of the HBI in galaxy clusters, but does not prevent it.
    \item Even if the HBI was inefficient, and the magnetic field remained tangled, the resulting effective conductivity of $\fsz = 1/3$ Spitzer is not sufficiently high to influence the AGN self-regulated cooling flow. 
    \item Only very high values of \fsz, which would require predominantly radial magnetic fields, transport sufficient thermal energy to reduce the cluster cooling flow.
\end{itemize}

Our set of idealised simulations have demonstrated the inability of AGN jets alone to re-randomise the magnetic field after it has been tangentially aligned by the HBI, and, hence, to restore significant thermal conduction on cluster scales. We have shown that thermal conduction does affect the morphology of the magnetic field, which in turn could have important consequences for where cosmic ray energy is deposited in the cluster (see \citealp{Beckmann2022}). How the magnetic field evolves depends strongly on how conduction is modelled, and on the initial conditions of the magnetic field. We defer a more detailed study of the impact of magnetic field morphology on the evolution of galaxy clusters to future work, together with a study of the contribution of other sources of turbulence beyond AGN jets, such as large-scale inflows, stirring by satellites and other substructure ~\citep{RuszkowskiOh2011,Bourne2017}. Other phenomena to be considered are anisotropic thermal pressure due to Braginskii viscosity~\citep{Kunz2011}.

\begin{acknowledgements}
RSB And YD designed the project, RSB, YD and AP developed the code, interpreted results and wrote the paper. RSB designed, executed and processed the suite of simulations.  FLP and VO contributed to discussion and interpretation of results. This work was supported by the ANR grant LYRICS (ANR-16-CE31-0011) and was granted access to the HPC resources of CINES under the allocations A0080406955 and A0100406955 made by GENCI. This work has made use of the Infinity Cluster hosted by Institut d’Astrophysique de Paris. We thank Stéphane Rouberol for smoothly running this cluster for us. Visualisations in this paper were produced using the \textsc{yt project} \citep{Turk2011} and \textsc{cartopy} \citep{Cartopy}.  
\end{acknowledgements}

%%%%%%%%%%%%%%%%%%%%%%%%%%%%%%%%%%%%%%%%%%%%%%%%%%



\bibliographystyle{aa}
\bibliography{author.bib} % if your bibtex file is called example.bib

\begin{thebibliography}{57}
\expandafter\ifx\csname natexlab\endcsname\relax\def\natexlab#1{#1}\fi

\bibitem[{Andreon {et~al.}(2017)Andreon, Wang, Trinchieri, Moretti, \&
  Serra}]{Andreon2017}
Andreon, S., Wang, J., Trinchieri, G., Moretti, A., \& Serra, A.~L. 2017, \aap,
  606, A24

\bibitem[{Barnes {et~al.}(2019)Barnes, Kannan, Vogelsberger, Pfrommer,
  Puchwein, Weinberger, Springel, Pakmor, Nelson, Marinacci, Pillepich, Torrey,
  \& Hernquist}]{Barnes2019}
Barnes, D.~J., Kannan, R., Vogelsberger, M., {et~al.} 2019, \mnras, 488, 3003

\bibitem[{Beckmann {et~al.}(2019)Beckmann, Dubois, Guillard, Salome, Olivares,
  Polles, Cadiou, Combes, Hamer, Lehnert, \& {Pineau des
  Forets}}]{Beckmann2019b}
Beckmann, R.~S., Dubois, Y., Guillard, P., {et~al.} 2019, \aap, 631, A60

\bibitem[{Beckmann {et~al.}(2022)Beckmann, Dubois, Pellisier, Olivares, Polles,
  Hahn, Guillard, \& Lehnert}]{Beckmann2022}
Beckmann, R.~S., Dubois, Y., Pellisier, A., {et~al.} 2022, Preprint
  [\eprint[arXiv]{2204.03629}]

\bibitem[{Berlok {et~al.}(2021)Berlok, Quataert, Pessah, \&
  Pfrommer}]{Berlok2021}
Berlok, T., Quataert, E., Pessah, M.~E., \& Pfrommer, C. 2021, \mnras, 504,
  3435

\bibitem[{Binney \& Tabor(1995)}]{Binney1995}
Binney, J. \& Tabor, G. 1995, \mnras, 276, 663

\bibitem[{Bogdanovi{\'{c}} {et~al.}(2009)Bogdanovi{\'{c}}, Reynolds, Balbus, \&
  Parrish}]{Bogdanovic2009}
Bogdanovi{\'{c}}, T., Reynolds, C.~S., Balbus, S.~A., \& Parrish, I.~J. 2009,
  \apj, 704, 211

\bibitem[{Bourne \& Sijacki(2017)}]{Bourne2017}
Bourne, M.~A. \& Sijacki, D. 2017, \mnras, 472, 4707

\bibitem[{Cowie \& McKee(1977)}]{Cowie1977}
Cowie, L.~L. \& McKee, C.~F. 1977, \apj, 211, 135

\bibitem[{Dashyan \& Dubois(2020)}]{Dashyan2020}
Dashyan, G. \& Dubois, Y. 2020, \aap, 638, A123

\bibitem[{Drake {et~al.}(2021)Drake, Pfrommer, Reynolds, Ruszkowski, Swisdak,
  Einarsson, Thomas, Hassam, \& Roberg-Clark}]{Drake2020}
Drake, J.~F., Pfrommer, C., Reynolds, C.~S., {et~al.} 2021, \apj, 923, 245

\bibitem[{Dubois {et~al.}(2021)Dubois, Beckmann, Bournaud, Choi, Devriendt,
  Jackson, Kaviraj, Kimm, Kraljic, Laigle, Martin, Park, Peirani, Pichon,
  Volonteri, \& Yi}]{Dubois2021}
Dubois, Y., Beckmann, R., Bournaud, F., {et~al.} 2021, \aap, 651, A109

\bibitem[{Dubois \& Commer{\c{c}}on(2016)}]{Dubois2016}
Dubois, Y. \& Commer{\c{c}}on, B. 2016, \aap, 585, A138

\bibitem[{Dubois {et~al.}(2012)Dubois, Devriendt, Slyz, \&
  Teyssier}]{Dubois2012}
Dubois, Y., Devriendt, J., Slyz, A., \& Teyssier, R. 2012, \mnras, 420, 2662

\bibitem[{Dubois {et~al.}(2014)Dubois, Volonteri, Silk, Devriendt, \&
  Slyz}]{Dubois2014spin}
Dubois, Y., Volonteri, M., Silk, J., Devriendt, J., \& Slyz, A. 2014, \mnras,
  440, 2333

\bibitem[{Fabian(1994)}]{Fabian1994}
Fabian, A.~C. 1994, \araa, 32, 277

\bibitem[{Fromang {et~al.}(2006)Fromang, Hennebelle, \& Teyssier}]{Fromang2006}
Fromang, S., Hennebelle, P., \& Teyssier, R. 2006, \aap, 457, 371

\bibitem[{Gaspari {et~al.}(2013)Gaspari, Ruszkowski, \& Oh}]{Gaspari2013}
Gaspari, M., Ruszkowski, M., \& Oh, S.~P. 2013, \mnras, 432, 3401

\bibitem[{Jacob \& Pfrommer(2017)}]{Jacob2017a}
Jacob, S. \& Pfrommer, C. 2017, \mnras, 467, stx131

\bibitem[{Kannan {et~al.}(2016)Kannan, Vogelsberger, Pfrommer, Weinberger,
  Springel, Hernquist, Puchwein, \& Pakmor}]{Kannan2017}
Kannan, R., Vogelsberger, M., Pfrommer, C., {et~al.} 2016, \apj, 837, L18

\bibitem[{{Karen Yang} \& Reynolds(2016)}]{Yang2016}
{Karen Yang}, H.-Y. \& Reynolds, C.~S. 2016, \apj, 818, 181

\bibitem[{Kimm {et~al.}(2015)Kimm, Cen, Devriendt, Dubois, \& Slyz}]{Kimm2015}
Kimm, T., Cen, R., Devriendt, J., Dubois, Y., \& Slyz, A. 2015, \mnras, 451,
  2900

\bibitem[{Komarov {et~al.}(2018)Komarov, Schekochihin, Churazov, \&
  Spitkovsky}]{Komarov18}
Komarov, S., Schekochihin, A., Churazov, E., \& Spitkovsky, A. 2018, J. Plasma
  Phys., 84, 905840305

\bibitem[{Kunz {et~al.}(2011)Kunz, Schekochihin, Cowley, Binney, \&
  Sanders}]{Kunz2011}
Kunz, M.~W., Schekochihin, A.~A., Cowley, S.~C., Binney, J.~J., \& Sanders,
  J.~S. 2011, \mnras, 410, 2446

\bibitem[{Leccardi \& Molendi(2008)}]{Leccardi2008}
Leccardi, A. \& Molendi, S. 2008, \aap, 487, 461

\bibitem[{Li {et~al.}(2015)Li, Bryan, Ruszkowski, Voit, O'Shea, \&
  Donahue}]{Li2015}
Li, Y., Bryan, G.~L., Ruszkowski, M., {et~al.} 2015, \apj, 811, 73

\bibitem[{Li {et~al.}(2017)Li, Ruszkowski, \& Bryan}]{Li2017}
Li, Y., Ruszkowski, M., \& Bryan, G. G.~L. 2017, \apj, 847, 106

\bibitem[{Maccio {et~al.}(2007)Maccio, Dutton, {Van Den Bosch}, Moore, Potter,
  \& Stadel}]{Maccio2007}
Maccio, A.~V., Dutton, A.~A., {Van Den Bosch}, F.~C., {et~al.} 2007, \mnras,
  378, 55

\bibitem[{McCourt {et~al.}(2011)McCourt, Parrish, Sharma, \&
  Quataert}]{McCourt2011}
McCourt, M., Parrish, I.~J., Sharma, P., \& Quataert, E. 2011, \mnras, 413,
  1295

\bibitem[{McKinney {et~al.}(2012)McKinney, Tchekhovskoy, \&
  Blandford}]{McKinney2012}
McKinney, J.~C., Tchekhovskoy, A., \& Blandford, R.~D. 2012, \mnras, 423, 3083

\bibitem[{{Met Office}(2015)}]{Cartopy}
{Met Office}. 2015, {Cartopy: a cartographic python library with a Matplotlib
  interface}

\bibitem[{Miyoshi \& Kusano(2005)}]{Miyoshi2005}
Miyoshi, T. \& Kusano, K. 2005, J. Comput. Phys., 208, 315

\bibitem[{Narayan \& Medvedev(2001)}]{Narayan01}
Narayan, R. \& Medvedev, M.~V. 2001, \apj, 562, L129

\bibitem[{Navarro {et~al.}(1997)Navarro, Frenk, \& White}]{Navarro1997}
Navarro, J.~F., Frenk, C.~S., \& White, S. D.~M. 1997, \apj, 490, 493

\bibitem[{Parrish {et~al.}(2012)Parrish, McCourt, Quataert, \&
  Sharma}]{Parrish2012}
Parrish, I.~J., McCourt, M., Quataert, E., \& Sharma, P. 2012, \mnras, 422, 704

\bibitem[{Parrish {et~al.}(2009)Parrish, Quataert, \& Sharma}]{Parrish2009}
Parrish, I.~J., Quataert, E., \& Sharma, P. 2009, \apj, 703, 96

\bibitem[{Phipps {et~al.}(2019)Phipps, Bogd{\'{a}}n, Lovisari, Kov{\'{a}}cs,
  Volonteri, \& Dubois}]{Phipps2019}
Phipps, F., Bogd{\'{a}}n, {\'{A}}., Lovisari, L., {et~al.} 2019, \apj, 875, 141

\bibitem[{Prasad {et~al.}(2015)Prasad, Sharma, \& Babul}]{Prasad2015a}
Prasad, D., Sharma, P., \& Babul, A. 2015, \apj, 811, 108

\bibitem[{Quataert(2008)}]{Quataert2008}
Quataert, E. 2008, \apj, 673, 758

\bibitem[{Roberg-Clark {et~al.}(2016)Roberg-Clark, Drake, Reynolds, \&
  Swisdak}]{Roberg2016}
Roberg-Clark, G.~T., Drake, J.~F., Reynolds, C.~S., \& Swisdak, M. 2016, \apj,
  830, L9

\bibitem[{Rosen \& Bregman(1995)}]{Rosen1995}
Rosen, A. \& Bregman, J.~N. 1995, \apj, 440, 634

\bibitem[{Ruszkowski \& Begelman(2002)}]{Ruszkowski02}
Ruszkowski, M. \& Begelman, M.~C. 2002, \apj, 581, 223

\bibitem[{Ruszkowski {et~al.}(2011)Ruszkowski, Lee, Br{\"{u}}ggen, Parrish, \&
  Oh}]{Ruszkowski2011}
Ruszkowski, M., Lee, D., Br{\"{u}}ggen, M., Parrish, I., \& Oh, S.~P. 2011,
  \apj, 740, 81

\bibitem[{Ruszkowski \& Oh(2010)}]{Ruszkowski2010}
Ruszkowski, M. \& Oh, S.~P. 2010, \apj, 713, 1332

\bibitem[{Ruszkowski \& Oh(2011)}]{RuszkowskiOh2011}
Ruszkowski, M. \& Oh, S.~P. 2011, \mnras, 414, 1493

\bibitem[{Santos(2000)}]{DosSantos2001}
Santos, S.~D. 2000, \mnras, 323, 930

\bibitem[{Sarazin(1986)}]{Sarazin1986}
Sarazin, C.~L. 1986, Rev. Mod. Phys., 58, 1

\bibitem[{Sharma {et~al.}(2007)Sharma, Quataert, Hammett, \&
  Stone}]{Sharma2007}
Sharma, P., Quataert, E., Hammett, G.~W., \& Stone, J.~M. 2007, \apj, 667, 714

\bibitem[{Spitzer \& H{\"{a}}rm(1953)}]{Spitzer1953}
Spitzer, L. \& H{\"{a}}rm, R. 1953, Phys. Rev., 89, 977

\bibitem[{Su {et~al.}(2019)Su, Hopkins, Hayward, Ma, Faucher-Gigu{\`{e}}re,
  Kere{\v{s}}, Orr, Chan, \& Robles}]{Su2019}
Su, K.-Y., Hopkins, P.~F., Hayward, C.~C., {et~al.} 2019, \mnras, 487, 4393

\bibitem[{Sutherland \& Dopita(1993)}]{Sutherland1993}
Sutherland, R.~S. \& Dopita, M.~A. 1993, Astrophys. J. Suppl. Ser., 88, 253

\bibitem[{Teyssier(2002)}]{Teyssier2002}
Teyssier, R. 2002, \aap, 385, 337

\bibitem[{Teyssier {et~al.}(2006)Teyssier, Fromang, \& Dormy}]{Teyssier2006}
Teyssier, R., Fromang, S., \& Dormy, E. 2006, J. Comput. Phys., 218, 44

\bibitem[{Tucker \& Rosner(1983)}]{Tucker1983}
Tucker, W.~H. \& Rosner, R. 1983, \apj, 267, 547

\bibitem[{Turk {et~al.}(2011)Turk, Smith, Oishi, Skory, Skillman, Abel, \&
  Norman}]{Turk2011}
Turk, M.~J., Smith, B.~D., Oishi, J.~S., {et~al.} 2011, Astrophys. J. Suppl.
  Ser., 192, 9

\bibitem[{Urban {et~al.}(2017)Urban, Werner, Allen, Simionescu, \&
  Mantz}]{Urban2017}
Urban, O., Werner, N., Allen, S.~W., Simionescu, A., \& Mantz, A. 2017, \mnras,
  470, 4583

\bibitem[{Wagh {et~al.}(2014)Wagh, Sharma, \& McCourt}]{Wagh2014}
Wagh, B., Sharma, P., \& McCourt, M. 2014, \mnras, 439, 2822

\end{thebibliography}


%%%%%%%%%%%%%%%%%%%%%%%%%%%%%%%%%%%%%%%%%%%%%%%%%%

%%%%%%%%%%%%%%%%% APPENDICES %%%%%%%%%%%%%%%%%%%%%

\begin{appendix}
%  \section{Magnetic field strength}
% \begin{figure}
%     \centering
%     \includegraphics[width=\columnwidth]{Figures/Bfield_evolution.png}
%     \caption{Magnetic field strength as a function of radius for the initial conditions (dotted), and at $t=1.5$ (dashed) and $3 \rm \ Gyr$ (solid).\rb{Yohan, what was it that you wanted to say about this?}}
%     \label{fig:bfield_evolution}
% \end{figure}

\end{appendix}

%%%%%%%%%%%%%%%%%%%%%%%%%%%%%%%%%%%%%%%%%%%%%%%%%%

\end{document}